\documentclass[3p]{elsarticle}
\usepackage{ALgo}
\usepackage{amsmath,amssymb}
\usepackage{url}
\usepackage{tikz, pgfplots}
\usepackage{multirow}

\usepackage[position=top]{subfig}

\def\sa#1{\mbox{\tt #1}}
\def\pp{\mathinner{\ldotp\ldotp}}
\def\PV{\mathcal{P}}
\def\PVW{\mathcal{P}_{\s{w}}}
\def\QV{\mathcal{Q}} 

\def\Div{\mathit{Div}}
\def\Mult{\mathit{Mult}}

\def\myCount{\textit{count}}

\newtheorem{example}{Example}

\newtheorem{thm}{Theorem}
\newtheorem{lem}[thm]{Lemma}
\newtheorem{prop}[thm]{Proposition}
\newdefinition{defi}{Definition}
\newdefinition{obs}{Observation}
\newproof{pf}{Proof}

\def\s#1{\mbox{\boldmath $#1$}}
\def\itbf#1{\textit{\textbf{#1}}}

\renewcommand{\epsilon}{\varepsilon}

\begin{document}

\title{\textbf{A Note on Easy and Efficient Computation \\of Full Abelian Periods of a Word}\footnote{The results in this note have been presented in preliminary form in \cite{FiLeLePrSm12}.}}

\author[palermo]{G. Fici}
\ead{Gabriele.Fici@unipa.it}

\author[rouen]{T. Lecroq}
\ead{Thierry.Lecroq@univ-rouen.fr}

\author[rouen]{A. Lefebvre}
\ead{Arnaud.Lefebvre@univ-rouen.fr}

\author[rouen]{\'E. Prieur-Gaston}
\ead{Elise.Prieur@univ-rouen.fr}

\author[mcmaster]{W.\ F.\ Smyth}
\ead{smyth@mcmaster.ca}

\address[palermo]{Dipartimento di Matematica e Informatica, Universit\`a di Palermo, Italy}

\address[rouen]{LITIS EA4108, Universit\'e de Rouen, 76821 Mont-Saint-Aignan Cedex, France}

\address[mcmaster]{Dept of Computing and Software, McMaster University,
Hamilton ON L8S 4K1, Canada and Faculty of Engineering \& Information Technology,
 Murdoch University, Murdoch WA 6150, Australia}

\begin{abstract}
Constantinescu and Ilie (Bulletin of the EATCS 89, 167--170, 2006) introduced the idea of an Abelian period with head and tail of a finite word. An Abelian period is called full if both the head and the tail are empty.
We present a simple and easy-to-implement $O(n\log\log n)$-time algorithm for computing all the full Abelian periods of a word of length $n$ over a constant-size alphabet. Experiments show that our algorithm significantly outperforms the $O(n)$ algorithm proposed by Kociumaka et al.~(Proc.~of STACS, 245--256, 2013) for the same problem.
\end{abstract}


\maketitle

\section{Introduction}

The study of repetitions in words is a classical topic in Stringology. A word is called an (integer) power if it can be written as the concatenation of two or more copies of another word, like \sa{barbar}. However, any word can be written as a \emph{fractional} power; that is, given a word $\s{w}$, one can always find a word $\s{u}$ such that $\s{w}=\s{u}^{n}\s{u'}$, where $\s{u'}$ is a (possible empty) prefix of $\s{u}$ and $n$ is an integer greater than or equal to one. In this case, the length of $\s{u}$ is called \emph{a period} of the word $\s{w}$. A word $\s{w}$ can have different periods, the least of which is usually called \emph{the period} of $\s{w}$.

Recently, a natural extension of this setting has been considered involving the notion of commutative equivalence. Two words are called commutatively equivalent if they have the same number of occurrences of each letter; that is, if one is an anagram of the other. An Abelian power (also called a weak repetition~\cite{Cummings_weakrepetitions}) is a word that can be written as the concatenation of two or more words that are commutatively equivalent, like  \sa{iceddice}.

Recall that the Parikh vector $\PVW$ of a word $\s{w}$ is the vector whose $i$th entry is the 
 number of occurrences of the $i$th letter of the alphabet in $\s{w}$. 
 For example, given the (ordered) alphabet $\Sigma=\{a,b,c\}$, the Parikh vector
 of the word $\s{w}=aaba$ is $\PVW=(3,1,0)$. Two words are therefore commutatively equivalent if and only if they have the same Parikh vector.

Constantinescu and Ilie~\cite{CI2006} introduced the definition of an Abelian period with head and tail of a word $\s{w}$ over a
 finite ordered alphabet $\Sigma=\{a_{1},a_{2},\ldots , a_{\sigma}\}$: An integer $p>0$ is an Abelian period of $\s{w}$  if one can write
 $\s{w}=\s{u}_0\s{u}_1 \cdots \s{u}_{k-1}\s{u}_k$ where for $0<i<k$
 all the factors $\s{u}_i$'s have the same Parikh vector $\PV$ such that
 $\sum_{j=1}^{\sigma}\PV[j]=p$
 and the Parikh vectors of $\s{u}_0$ and $\s{u}_k$ are ``contained''
 in $\PV$, in the sense that they are proper sub-Parikh vectors of $\PV$ (see next section for the formal definition of ``contained''). In this case, $\s{u}_0$ and $\s{u}_k$ are called the head and the tail of the Abelian period $p$, respectively. 
This definition of an Abelian period matches that of an
Abelian power when $\s{u}_0$ and $\s{u}_k$ are both  empty  and
 $k>2$.
 
 As an example, the word $\s{w}=abaababa$ over the alphabet $\Sigma=\{a,b\}$ can be written as
 $\s{w}=\s{u}_{0}\s{u}_{1}\s{u}_{2}\s{u}_{3}$, where
 $\s{u}_{0}=ab$, $\s{u}_{1}=aab$, $\s{u}_{2}=aba$, $\s{u}_{3}=\varepsilon$, with $\varepsilon$ the empty word,
 so that $3$ is an Abelian period of $\s{w}$ with Parikh vector $(2,1)$
 (the Parikh vector of $\s{u}_{0}$ is $(1,1)$ and that of $\s{u}_{3}$
 is $(0,0)$ which are both ``contained'' in $(2,1)$). Notice that $\s{w}$ has also Abelian period $2$, since it can be written as $\s{w}=\s{u}_{0}\s{u}_{1}\s{u}_{2}\s{u}_{3}\s{u}_{4}$, with $\s{u}_{0}=a$, $\s{u}_{1}=ba$, $\s{u}_{2}=ab$, $\s{u}_{3}=ab$, $\s{u}_{4}=a$.

This example shows that a word can have different Abelian periods. Moreover, a word can have the same Abelian period $p$ corresponding to different factorizations; that is, with different heads. Actually, a word of length $n$ can have $\Theta(n^{2})$ many different Abelian periods \cite{DAM}, if these are represented in the form $(h,p)$, where $h$ is the length of the head---the length of the tail is uniquely determined by $h$ and $p$.

Recently~\cite{PSC2011-16,DAM} we described algorithms for computing all the
 Abelian periods
 of a word of length $n$ in time $O(n^2 \times \sigma)$.
This was improved to time $O(n^2)$ in~\cite{MichalisChrist2012}.
In~\cite{CIKKPRRTW2012} the authors derived an 
 efficient
 algorithm for computing the Abelian periods 
based on prior computation of the Abelian squares.

An Abelian period is called \emph{full} if both the head and the tail are empty. Clearly, a full Abelian period is a divisor of the length of the word.

A preliminary version of the present paper appeared in~\cite{FiLeLePrSm12}
 where we presented brute force algorithms to compute full Abelian periods
 and Abelian periods without head and with tail in $O(n^2)$ time and a 
 quasi-linear time algorithm QLFAP for
 computing all the full Abelian periods of a word.
In \cite{KoRaRy13} Kociumaka et al.~gave a linear time algorithm LFAP for the same problem.
Here we first briefly outline LFAP,
followed by a description of QLFAP.
Then, extending the presentation in~\cite{FiLeLePrSm12},
we add an experimental section to demonstrate that our algorithm
significantly outperforms LFAP in practice,
both on pseudo-randomly generated and genomic data.
Our method has the additional
advantage of being conceptually simple and easy to implement.
  
\section{Notation}
\label{sec-def}

Let $\Sigma=\{a_{1},a_{2},\ldots,a_{\sigma}\}$ be a finite ordered
 alphabet of cardinality $\sigma$ and $\Sigma^*$ the set of finite words
 over $\Sigma$. 
We let $|\s{w}|$ denote the length of the word $\s{w}$.
Given a word $\s{w}=\s{w}[0\pp n-1]$ of length $n>0$, we write $\s{w}[i]$ for the $(i+1)$-th symbol of $\s{w}$
 and, for $0\leqslant i \leqslant j< n$, we write $\s{w}[i\pp j]$ for the factor of $\s{w}$
 from the $(i+1)$-th symbol to the $(j+1)$-th symbol, both included.
We let $|\s{w}|_a$ denote the number of occurrences of the symbol
 $a\in\Sigma$ in the word $\s{w}$. 

The \emph{Parikh vector} of $\s{w}$, denoted by $\PVW$,
 counts the
 occurrences of each letter of $\Sigma$ in $\s{w}$, that is, 
 $\PVW=(|\s{w}|_{a_{1}},\ldots,|\s{w}|_{a_{\sigma}})$.
Notice that two words have the same Parikh vector if and only if
 one word is a permutation of the other (in other words, an anagram).

Given the Parikh vector $\PVW$ of a word $\s{w}$, we let $\PVW [i]$ denote its
 $i$-th component and $|\PVW|$ its norm, defined as the sum of its components.
Thus, for $\s{w}\in\Sigma^*$ and $1\leqslant i\leqslant\sigma$, we have
 $\PVW [i]=|\s{w}|_{a_i}$ and $|\PVW|=\sum_{i=1}^{\sigma}\PVW[i]=|\s{w}|$.

Finally, given two Parikh vectors $\PV,\QV$, we write $\PV\subset \QV$ if
 $\PV[i]\leqslant \QV[i]$
 for every $1\leqslant i\leqslant \sigma$ and $|\PV|<|\QV|$. 
This makes precise the notion of ``contained'' used in the Introduction.
 
\begin{defi}[Abelian period \cite{CI2006}]
\label{def-ap}
A word $\s{w}$ has an Abelian period $(h,p)$ if
 $\s{w}=\s{u}_0\s{u}_1 \cdots \s{u}_{k-1}\s{u}_{k}$ 
 such that:

\begin{itemize}
 \item $\PV_{\s{u}_{0}}\subset \PV_{\s{u}_{1}}=\cdots =\PV_{\s{u}_{k-1}}\supset \PV_{\s{u}_{k}}$,
 \item $|\s{u}_{0}|=h$, $|\s{u}_{1}|=p$.
\end{itemize}

\end{defi}

We call $\s{u}_0$ and $\s{u}_k$ respectively the \emph{head} and the
 \emph{tail}  of the Abelian period.
 Notice that the length $t=|\s{u}_k|$ of the tail is uniquely determined
 by $h$, $p$ and $|\s{w}|$, namely $t=(|\s{w}|-h) \bmod p$. 

The following lemma gives a bound on the maximum number of Abelian periods
 of a word.
\begin{lem}[\cite{PSC2011-16}]
\label{lemma-max}
The maximum number of different Abelian periods $(h,p)$ for a word of length $n$ over
 an alphabet of size $\sigma$
 is $\Theta(n^2)$.
\end{lem}

\begin{pf}
The word $(a_{1}a_{2}\cdots a_{\sigma})^{n/\sigma}$ has Abelian period
 $(h,p)$ for any $p\equiv 0 \bmod \sigma$ and every $h$ such that 
 $0 \leqslant h \leqslant \min(p-1,n-p)$. \qed
\end{pf}

An Abelian period is called \emph{full} if it has head and tail both empty.
We are interested in computing all the full Abelian periods of a word. Notice that a full Abelian period of a word of length $n$ is a divisor of $n$.
In the remainder of this note, we will therefore write that a word $\s{w}$
 has an Abelian period $p$ if and only if it has full Abelian period $(0,p)$. 
 
\section{Previous work}\label{sec-prev-work}

We now outline the linear algorithm LFAP given in~\cite{KoRaRy13}.

Let $w$ be a word of length $n$.
Let $\PV_{w_i}=\PV_{w[0\pp i]}$.
Two positions $i,j\in \{1,\ldots,n\}$ are called proportional,
 which is denoted by $i\sim j$, if $\PV_{w_i}[k] = c\times \PV_{w_j}[k]$
 for each $k$, where $c$ is a real number independent of $k$.
 
An integer $k$ is called a candidate (as a potential Abelian period) if
 $i\sim k$ for each $i\in \Mult(k,n)$ where $\Mult(k,n)$ is the set
  of multiples of $k$ not exceeding $n$, or equivalently
 $k\sim 2k \sim 3k \sim \cdots$.

A positive integer $q \le n$ is a full Abelian period of $w$
 if and only if $q\mid n$ and $q$ is a candidate.

The algorithm first computes the set $A=\{ k : k\sim n\}$ and
 then the set
 $B = \{d \in \Div(n) : \Mult(d, n) \subseteq A\}$.

Let $\ell$ be the rank in the alphabet of a least frequent letter $a$ of $w$.
Let $q_0$ be the position of the first occurrence of $a$ in $w$.
For $i \in \{q_0, \ldots , n\}$ let 
 $\gamma_i = \PV_{w_i}/\PV_{w_i}[\ell]$.
Vectors $\gamma_i$ are used in order to deal with vector equality
 instead of vector proportionality.

If $i, j \in \{q_0, \ldots , n\}$ then $i \sim j$ is equivalent to 
 $\gamma_i = \gamma_j$.
The problem of computing the set $A$
 reduces to check if $\delta_i=\gamma_n-\gamma_i$
 is equal to $(0,\ldots,0)$ for $q_0\le i \le n-1$.
This is done in linear time by using what the authors called
 a diff-representation of the $\gamma_i$'s (see~\cite{KoRaRy13}
 for details).

The authors show that the set $B$ can be computed
 in linear time.
This can be done by showing that after an $O(n)$ time preprocessing,
 the value $\gcd(k,\ell)$ for any $k,\ell\in\{1,\ldots,n\}$ can be computed
 in constant time (see~\cite{KoRaRy13}
 for details).

\section{The new algorithm}\label{sec-nohead-notail}

In this section, we describe our algorithm QLFAP for computing all the full Abelian periods of a word $\s{w}$ of length $n$ over a constant-size alphabet $\Sigma=\{a_1,a_2,\ldots,a_\sigma\}$.

In a linear-time preprocessing phase, we compute $\PVW[j]$ for $j = 1,2,\ldots,\sigma$, 
 i.e., the components of the Parikh vector of the word $\s{w}$.
Let $g$ denote the greatest common divisor of the
 elements of the Parikh vector of $\s{w}$, computable in
 $O(\sigma + \log n/\sigma)$ time~\cite{Bradley1970}.
Then we compute $s = n/g$.
We can suppose $\sigma \ge 2$ and $g > 1$, otherwise the solution is trivial.
In $O(\sqrt{g})$ time we compute a stack $D$ of all divisors $d \ge 1$ of $g$
 in increasing order. 

\begin{obs}
\label{obs-poss}
The only possible full Abelian periods of $\s{w}$
 are of the form $p = d\times s$, where $d$ is an entry in $D$.
Therefore, the smallest possible full Abelian period of $\s{w}$ is $s$.
\end{obs}

\begin{defi}[scaled]
\label{def-scaled}
A factor $\s{u}$ of a word $\s{w}$ is \itbf{scaled} if and only if there exists $k\geqslant 0$ such that $\PV_{\s{u}} = k \times \PVW / s$, where $s=n/g$ is
 the smallest possible Abelian period of $\s{w}$.
\end{defi}

\begin{obs}
\label{obs-scaled}
If $\s{w}=\s{uv}$ and $\s{u}$ is scaled, then $\s{v}$ is scaled.
\end{obs}

A scaled factor is called \emph{irreducible} if it cannot itself be factored into scaled factors. According to Definition~\ref{def-scaled} and 
 Observation~\ref{obs-scaled}, 
 every word $\s{w}$
 can be factored uniquely into irreducible scaled factors by computing the shortest scaled 
 prefix $\s{u}$ of $\s{w}=\s{uv}$ and then factoring the suffix $\s{v}$ recursively, 
 until this suffix is empty. This factorization is computed by calling procedure
\textsc{ComputeL} given in \figurename~\ref{algo-computel}, which computes the scaled prefixes of $\s{w}$ (or analogously, by the previous observation, the scaled suffixes of $\s{w}$). 
It returns a boolean array $L$ of length $n$ defined by: $L[i]=1$ if and only if $i$ is the starting position of a scaled suffix of $\s{w}$. 
It also returns the value $T$ such that the longest scaled factor computed is of length $s\times T$.

\begin{obs}
\label{obs-fact}
Consider two positions $i,j$, with $0\leqslant i<j<|\s{w}|$, such that $L[i]=L[j]=1$. Then $\s{w}[i\pp j-1]$ is scaled.
\end{obs}

\begin{obs}
\label{obs-T}
A full Abelian period of $w$ must be greater than or equal to $s\times T$.
\end{obs}

\begin{figure}
\begin{algo}[numcom]{ComputeL}{\s{w},s,g,\PVW}
  \SET{(i,T,L)}{(0,0,0^{|\s{w}|})}
  \DOWHILE{i\leqslant|\s{w}|-s}
    \SET{j}{0}
    \SET{t}{0}
    \SET{\myCount}{0^{\sigma}}
    \label{loop-scaled}
    \DOWHILE{j<\sigma}
      \INCR{t}
      \label{loop-PV}
      \DOFORI{k}{1}{s}
        \INCR{\myCount[w[i]]}
        \INCR{i}
      \OD
      \SET{j}{0}
      \label{internalloop}
      \DOWHILE{j<\sigma \AND \lfloor \myCount[j]/t\rfloor=\PVW[j]/g}
        \INCR{j}
      \OD
    \OD
    \SET{L[i-t\times s]}{1}
    \SET{T}{\max(T,t)}
  \OD  
  \RETURN{(L,T)}
\end{algo}
\caption{\label{algo-computel}Algorithm computing 
array $L$ such that $L[i]=1$ iff $i$ is the starting position of a scaled
suffix of $\s{w}$.
}
\end{figure}

\begin{prop}
The algorithm \textsc{ComputeL} computes the boolean array $L$ of a word $\s{w}$
 of length $n$
 over an alphabet of size $\sigma$ in time $O(n)$.
\end{prop}

\begin{pf}
The internal loop in line~\ref{loop-scaled} is performed until a scaled factor
 of length $t\times s$ is found: the loop in line~\ref{loop-PV} computes the Parikh vector of
 the factor $w[i-t\times s\ldotp\ldotp i]$ and the loop in line~\ref{internalloop} verifies that this
 factor is scaled: its length is equal to $t\times s$ and it should hold for every letter $j$
 that $\myCount[j]/(t\times s)=\PVW[j]/n$.
This can be rewritten as $\myCount[j]/t=\PVW[j]/g$ since $n=s\times g$.
Since, by definition of $g$, $\PVW[j]/g$ is an integer the test can be reduced to
 $\lfloor \myCount[j]/t \rfloor=\PVW[j]/g$ since it should hold for every letter.
This avoids to use real numbers.
Furthermore the $\PVW[j]/g$ are constant for a given $\s{w}$ and can thus be 
 precomputed.
When such a scaled factor is found, its starting position is $i-t\times s$
 and $L[i-t\times s]$ is set to $1$. Since the algorithm starts from 
 position $i=0$, therefore, according to Observation~\ref{obs-scaled}, array $L$ is filled correctly.

The algorithm visits each position $i$ in $\s{w}$ exactly once,
 and corresponding to each $i$ performs a constant-time processing.
The internal loop in line~\ref{internalloop} is performed at most $\sigma$
 times, every $s$ positions.
It is performed $|\s{w}|/s$ times and since $\sigma\leqslant s$, each
 iteration costs $O(s)$ time.
Thus the algorithm runs in time $O(|\s{w}|)$.\qed
\end{pf}

\begin{example}
$\s{w}=\sa{abaababbbabaabbabbaaabbababbaa}$, $\PVW=(15,15)$.
\begin{center}
\scalebox{0.775}{
\setlength{\tabcolsep}{.09cm}
\begin{tabular}{|l|cccccccccccccccccccccccccccccc|}
\hline
$i$&$0$&$1$&$2$&$3$&$4$&$5$&$6$&$7$&$8$&$9$&$10$&$11$&$12$&$13$&$14$&$15$&$16$&$17$&$18$&$19$&$20$&$21$&$22$&$23$&$24$&$25$&$26$&$27$&$28$&$29$\\
\hline
$\s{w}[i]$&\sa{a}&\sa{b}&\sa{a}&\sa{a}&\sa{b}&\sa{a}&\sa{b}&\sa{b}&\sa{b}&\sa{a}&
\sa{b}&\sa{a}&\sa{a}&\sa{b}&\sa{b}&\sa{a}&\sa{b}&\sa{b}&\sa{a}&\sa{a}&
\sa{a}&\sa{b}&\sa{b}&\sa{a}&\sa{b}&\sa{a}&\sa{b}&\sa{b}&\sa{a}&\sa{a}\\
\hline
$L[i]$&$1$&$0$&$1$&$0$&$0$&$0$&$0$&$0$&$1$&$0$&$1$&$0$&$1$&$0$&$1$&$0$&$1$&$0$&$0$&$0$&$1$&$0$&$1$&$0$&$1$&$0$&$1$&$0$&$0$&$0$\\
\hline
\end{tabular}
}
\end{center}
\noindent
The word $\s{w}$ is factored into irreducible scaled factors in the following way:
$\s{w}=\sa{ab}\cdot\sa{aababb}\cdot\sa{ba}\cdot\sa{ba}\cdot\sa{ab}\cdot\sa{ba}\cdot\sa{bbaa}\cdot\sa{ab}\cdot\sa{ba}\cdot\sa{ba}\cdot\sa{bbaa}$. Since the length of the longest word in this factorization is $6$ and $s=2$, the
value of $T$ is $3$.
\end{example}

\begin{lem}
\label{lem-seg}
The word $\s{w}$ has full Abelian period $p$ 
 if and only if for every
 $k = 0, 1, \ldots, |\s{w}|/p-1$, $k\times p$ is the starting position
 of a scaled suffix of $\s{w}$.
 \end{lem}

\begin{pf}
Suppose that $p$ 
is an Abelian period of $\s{w}$.
According to Observation~\ref{obs-fact},  
we have that for every
 $k = 0, 1, \ldots, |\s{w}|/p-1$ the position
 $k\times p$ is the starting position
 of a scaled factor and then of a scaled suffix of $\s{w}$. 
 
Conversely, if $k\times p$ is the starting position
 of a scaled suffix of $\s{w}$ for every $k = 0, 1, \ldots, |\s{w}|/p-1$, it follows from Observation~\ref{obs-fact} that $\s{w}[k\times p\pp (k+1)\times p-1]$ is scaled and consequently  
 that $p$ is an Abelian period of $\s{w}$.\qed
\end{pf}

The algorithm \CALL{QuasiLinearFullAbelianPeriods}{\s{w}},
 given in \figurename~\ref{alg-Arep}, computes all the full
 Abelian periods 
 of a word $\s{w}$.
It first computes the greatest common divisor $g$ of the elements of
 $\PVW$ (line~\ref{inst1}).
Then it computes $s=n/g$ (line~\ref{inst2}).
This is followed by the computation of the array $L$ giving the starting positions of the scaled
 suffixes and of the threshold $T$ (line~\ref{inst3}). 
Then it computes the stack $D$ of all the divisors of $g$ greater than or equal to $T$
 (line~\ref{inst4}).
Then it checks for every divisor $d$ of $g$ in $D$, whether all the multiples
 of $d$ are starting positions of scaled suffixes  (lines~\ref{inst5}--\ref{inst6}).

\begin{example}
$\s{w}=\sa{abaababbbabaabbabbaaabbababbaa}$. We have
$n=30$, $\PVW[1]=\PVW[2]=15$, $g=15$, $s=2$, $D=(1,3,5,15)$.
When $d=1$, $p=2$ and $4=2\times p$ is not a starting position of a scaled suffix.
When $d=3$, $p=6$ and $6$ is not a starting position of a scaled suffix.
When $d=5$, $p=10$ and $10$, $20$ are starting positions of 
 scaled suffixes, hence $10$ is a full Abelian period of $\s{w}$.
The case where $d=15$ is trivial since it corresponds to Abelian period $n$.
Thus the algorithm returns $\{10,30\}$, which is the set of full Abelian periods of $\s{w}$.
\end{example}

\begin{thm}
The algorithm \CALL{QuasiLinearFullAbelianPeriods}{\s{w}} computes all the full
 Abelian periods of a word $\s{w}$ of length $n$ 
over an alphabet of size $\sigma$  in time $O(n\log\log n)$ and space $O(n)$.
\end{thm}

\begin{pf}
The correctness comes from Observation~\ref{obs-poss} 
 and from Lemma~\ref{lem-seg}.
The algorithm \CALL{QuasiLinearFullAbelianPeriods}{\s{w}}
 scans all the multiples of
 the divisors $d \in D$, whose number is equal to the sum of the
 divisors of $s$, which in turn is $O(n\log\log n)$~\cite{gronwall}.
The algorithm requires $O(n)$ space for $L$ and $D$.
\qed
 \end{pf}

\begin{figure}
\begin{algo}[numcom]{QuasiLinearFullAbelianPeriods}{\s{w}=[\s{w}_0\ldotp\ldotp \s{w}_{n-1}]}
  \label{inst1}\SET{g}{\text{greatest common divisor of elements of }\PVW}
  \label{inst2}\SET{s}{n/g}
  \label{inst3}\SET{(L,T)}{\CALL{ComputeL}{\s{w},s,g,\PVW}}
  \label{inst4}\SET{D}{\text{divisors of $g$ greater or equal than $T$}}
  \SET{R}{\emptyset}
  \label{inst5}\DOWHILE{d<g}
    \SET{k}{1}
    \DOWHILE{k\times d\times s<n \AND L[k\times d\times s]=1}
       \INCR{k}
    \OD 
    \IF{n\leqslant k\times d\times s}
      \SET{R}{R\cup \{i\}}
    \FI
    \label{inst6}\SET{d}{\CALL{Pop}{D}}
  \OD
  \SET{R}{R\cup \{n\}}
  \RETURN{R}
\end{algo}
\caption{\label{alg-Arep}Algorithm computing the set of Abelian periods of word $\s{w}$.}
\end{figure}

\section{Experimental results}\label{sec:res}

\subsection{Implementation}

We implemented both \textsc{QuasiLinearFullAbelianPeriods} (QLFAP)
and the algorithm \textsc{LinearFullAbelianPeriods} (LFAP) of~\cite{KoRaRy13}
in a homogeneous way in \texttt{C}.

LFAP requires seven integer arrays of space $O(n)$ to compute
the diff-representation plus five more for the preprocessing of the set B
(see Sect.~\ref{sec-prev-work} for terminology).
On the other hand, QLFAP requires only
two integer arrays $L$ and $R$ of length $n$,
together with the Parikh vector of length $\sigma$
and the stack $D$ (implemented as an integer array of length $g$).

In both algorithms, the memory space for the different arrays has been
 allocated one by one.

All the experiments have been performed on a computer with a $1.3$ GHz Intel Core i5 processor and 4 Go 1600 MHz DDR3 RAM.

\subsection{Results}

\begin{figure}

  \subfloat[]{\label{fig:2:5}
\includegraphics[width=5.5cm,angle=-90]{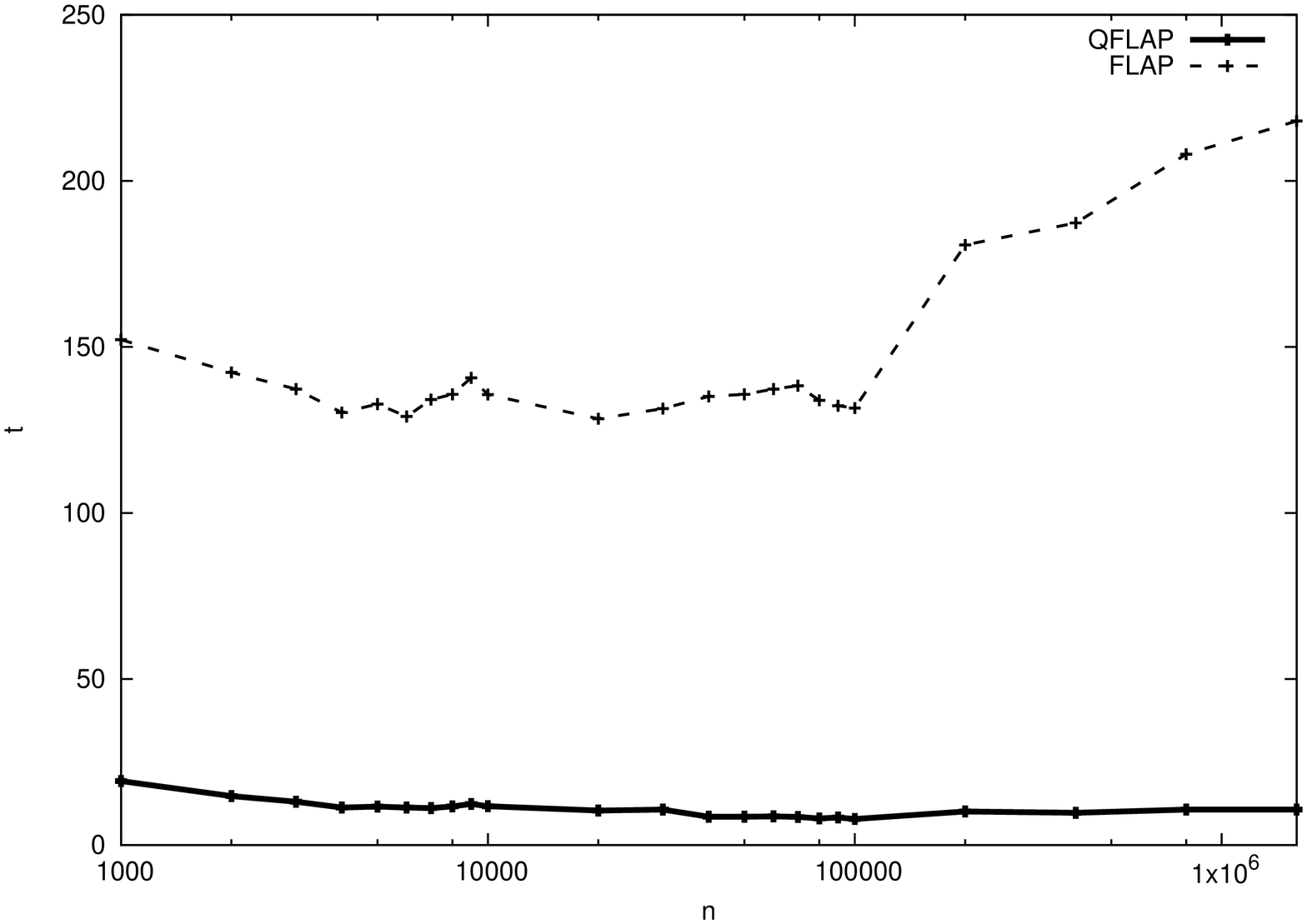}
  }\hspace{.3cm}
  \subfloat[]{\label{fig:5:5}
\includegraphics[width=5.5cm,angle=-90]{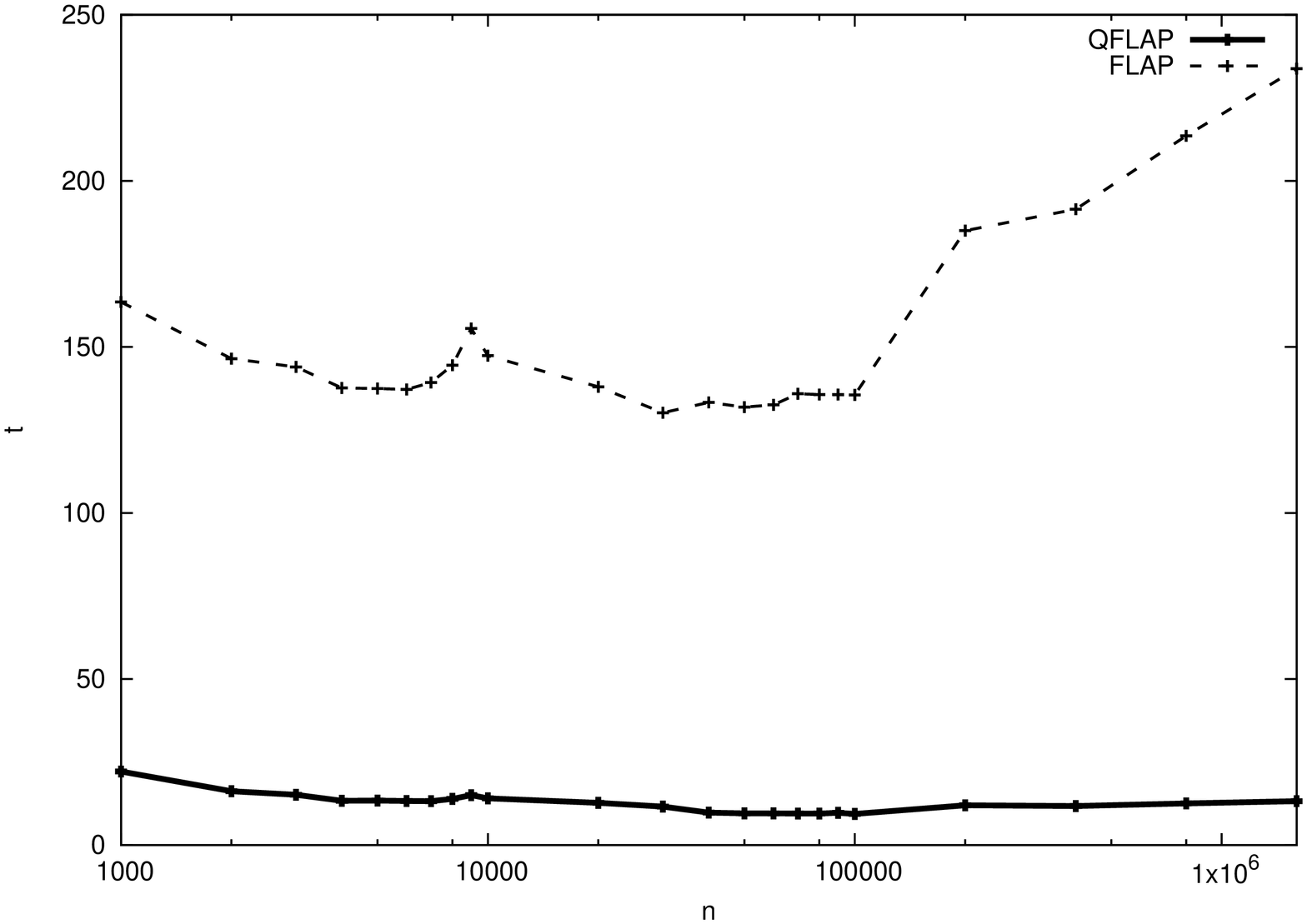}
}

  \subfloat[]{\label{fig:10:5}
\includegraphics[width=5.5cm,angle=-90]{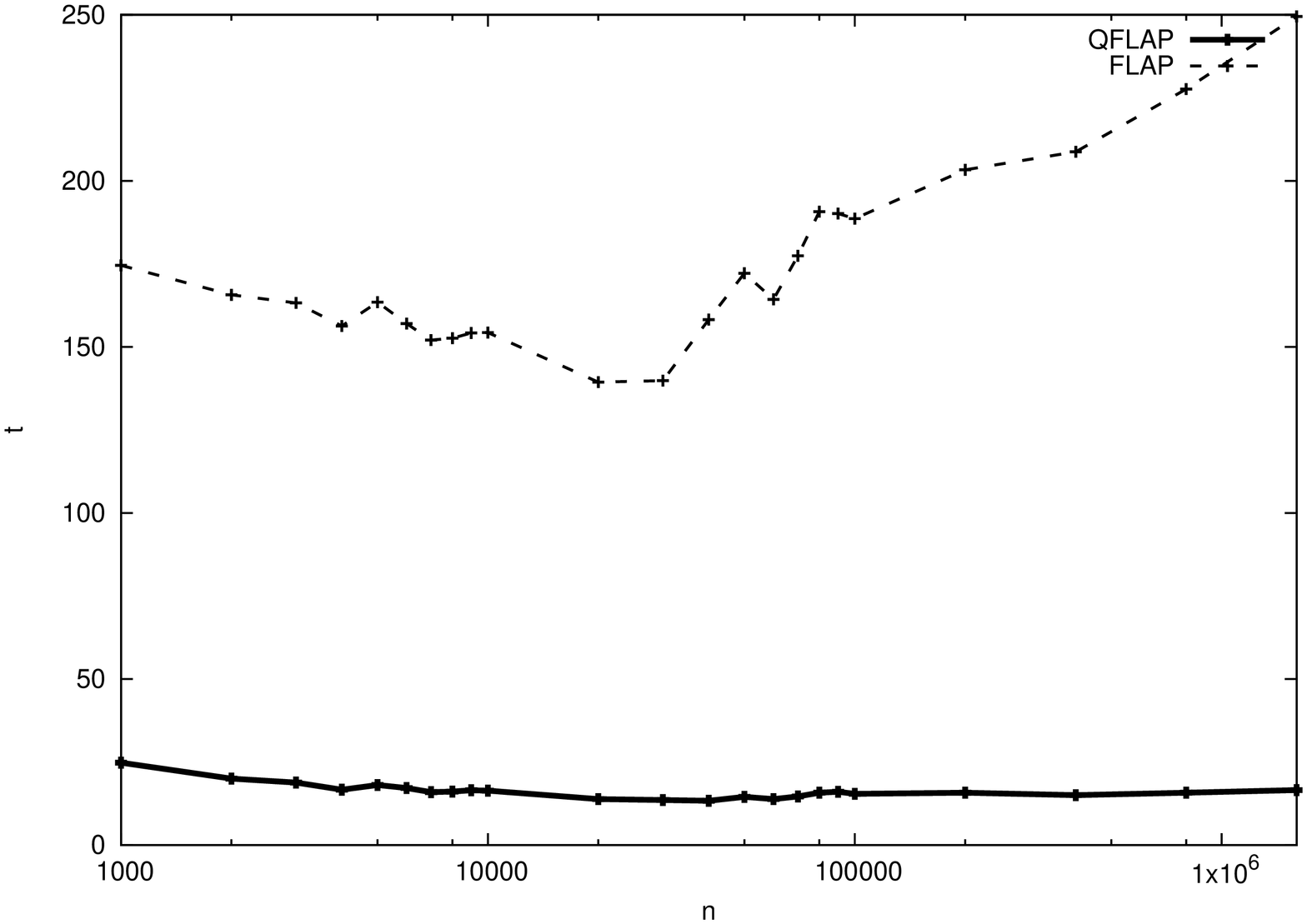}
  }\hspace{.3cm}
  \subfloat[]{\label{fig:20:5}
\includegraphics[width=5.5cm,angle=-90]{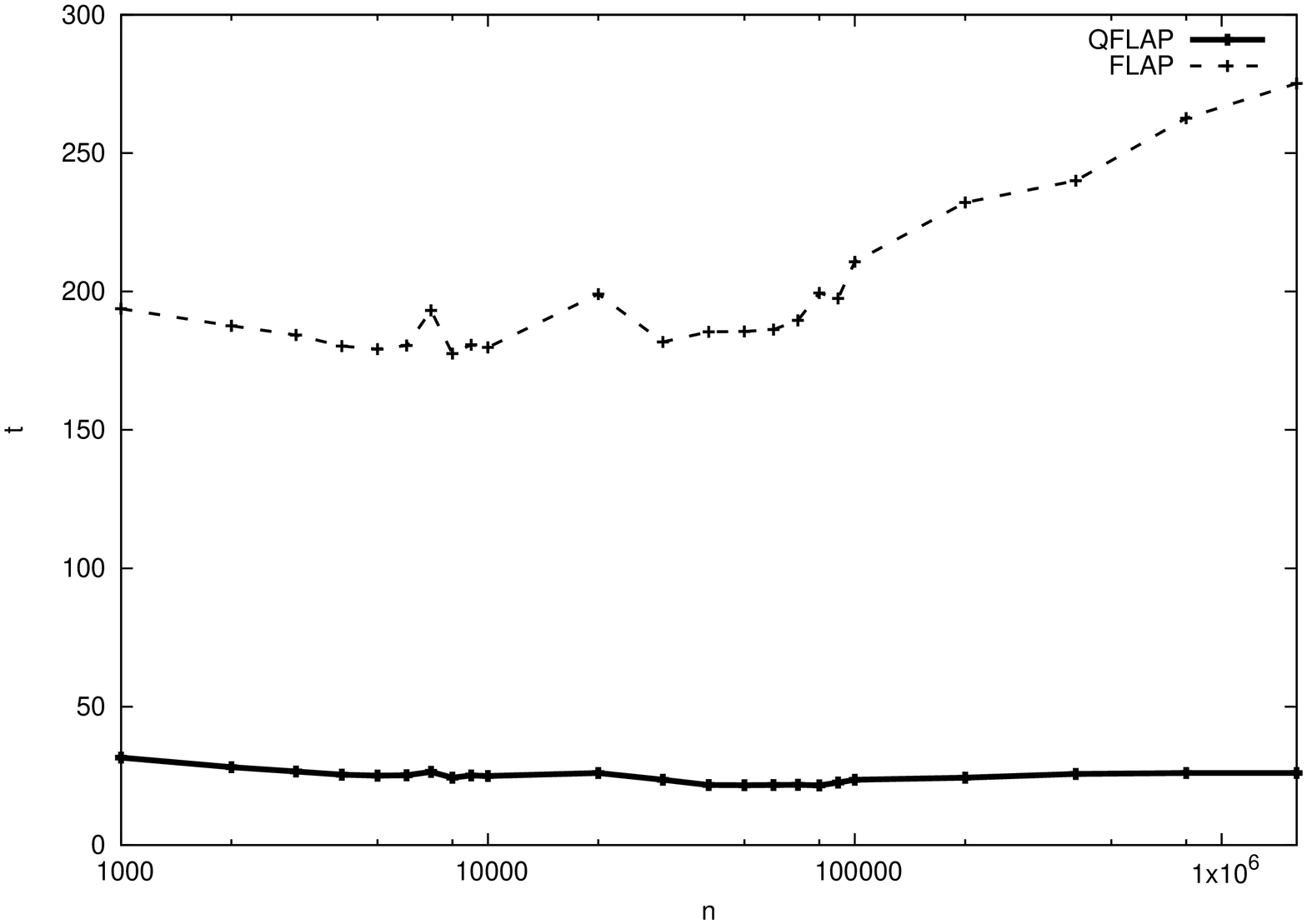}
}
\caption{
\label{fig-expe1}
Running times of the LFAP and QLFAP algorithms for
 \protect\subref{fig:2:5} alphabet of size 2 and at least one non trivial Abelian period 5; 
 \protect\subref{fig:5:5} alphabet of size 5 and at least one non trivial Abelian period 5; 
 \protect\subref{fig:10:5} alphabet of size 10 and at least one non trivial Abelian period 5; 
 \protect\subref{fig:20:5} alphabet of size 20 and at least one non trivial Abelian period 5.
}
\end{figure}

\begin{figure}
  \subfloat[]{\label{fig:2:20}
\includegraphics[width=5.5cm,angle=-90]{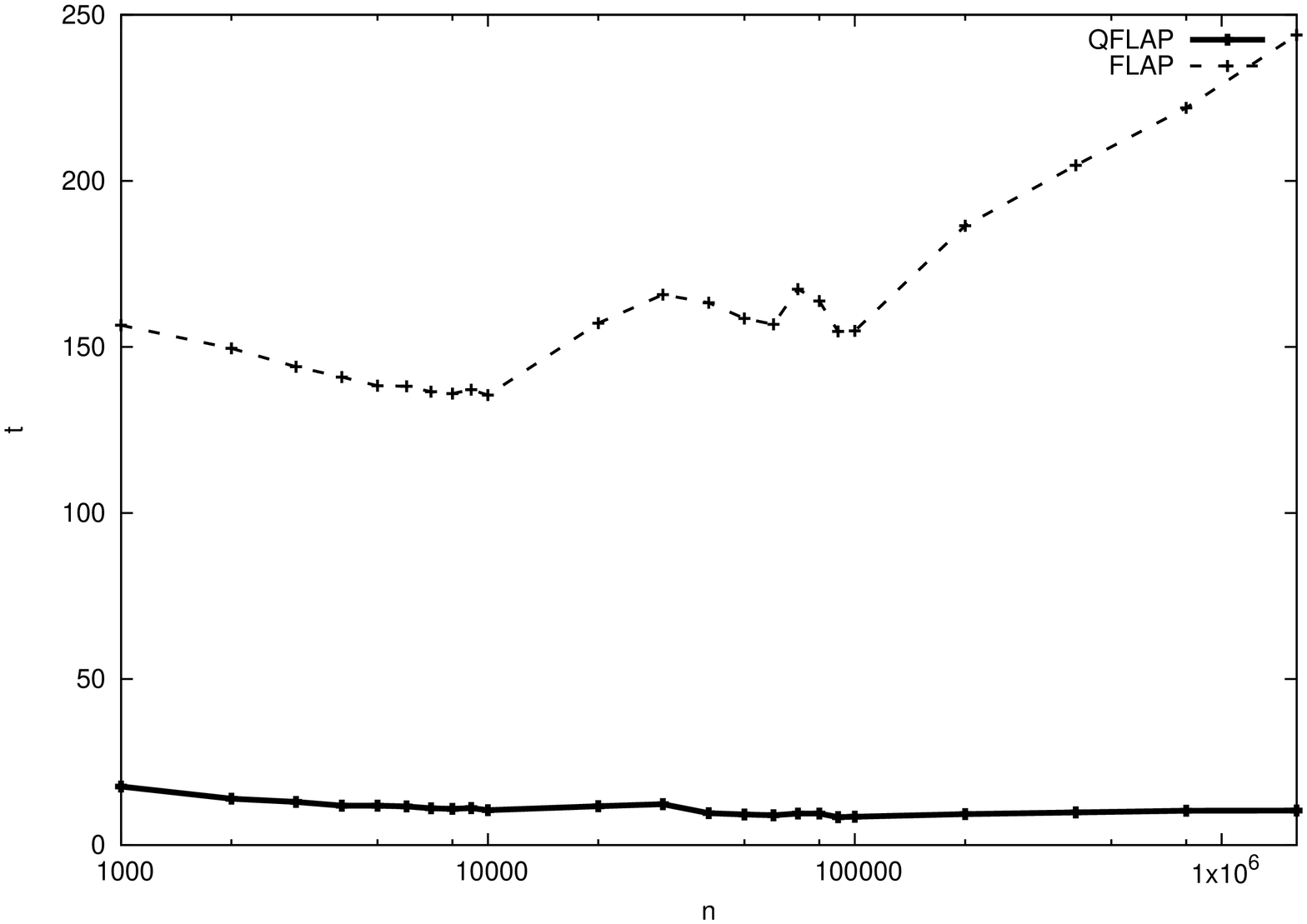}
  }\hspace{.3cm}
  \subfloat[]{\label{fig:5:20}
\includegraphics[width=5.5cm,angle=-90]{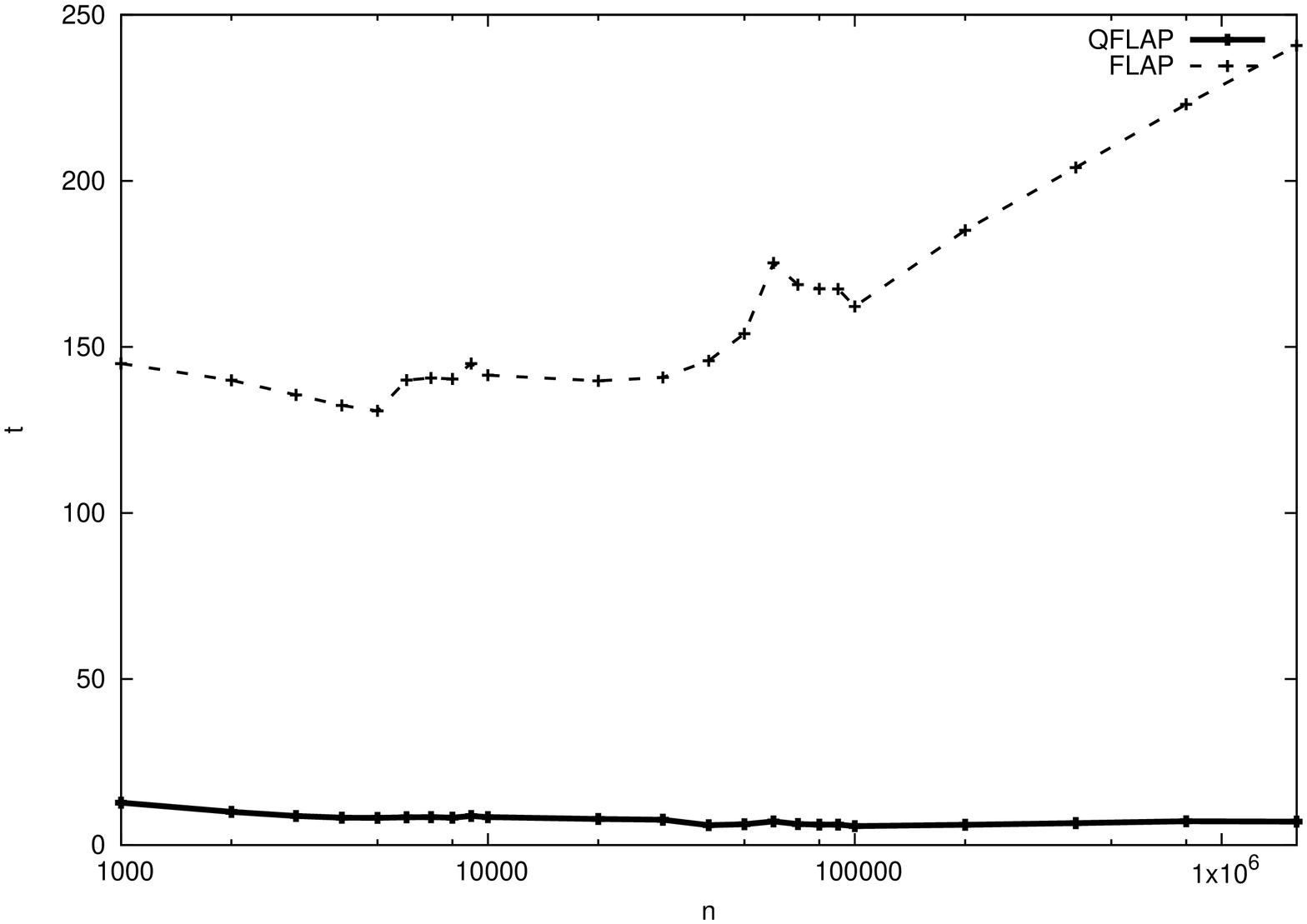}
}

  \subfloat[]{\label{fig:10:20}
\includegraphics[width=5.5cm,angle=-90]{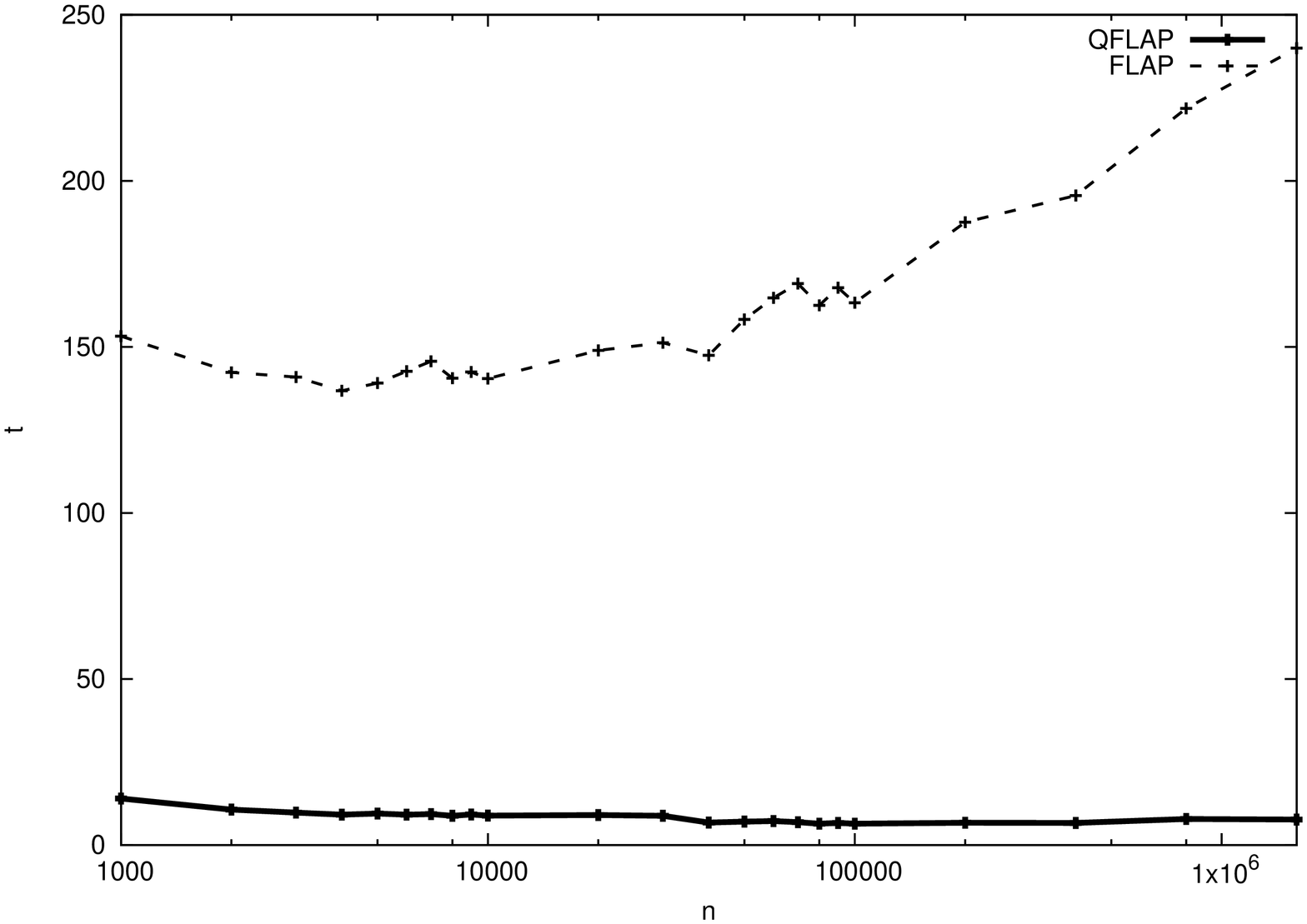}
  }\hspace{.3cm}
  \subfloat[]{\label{fig:20:20}
\includegraphics[width=5.5cm,angle=-90]{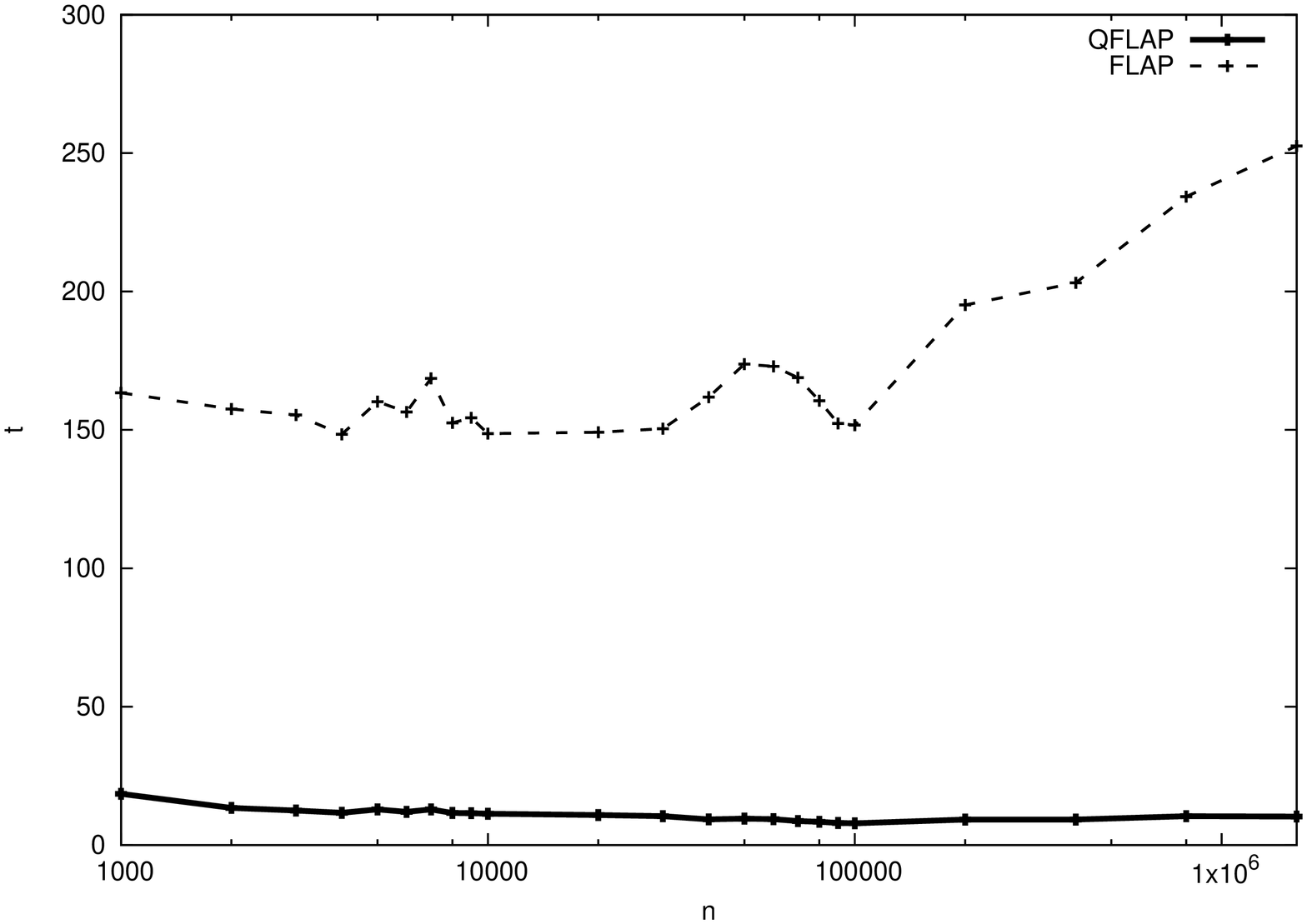}
}

\caption{
\label{fig-expe2}
Running times of the LFAP and QLFAP algorithms for
 \protect\subref{fig:2:20} alphabet of size 2 and at least one non trivial Abelian period 20; 
 \protect\subref{fig:5:20} alphabet of size 5 and at least one non trivial Abelian period 20; 
 \protect\subref{fig:10:20} alphabet of size 10 and at least one non trivial Abelian period 20; 
 \protect\subref{fig:20:20} alphabet of size 20 and at least one non trivial Abelian period 20.
}
\end{figure}

\figurename s~\ref{fig-expe1} and~\ref{fig-expe2} plots the raw execution times 
 versus the word length (in logarithmic scale)
while 
\figurename s~\ref{fig-ratio1}--\ref{fig-ratio3} shows execution time ratios
for LFAP over QLFAP.
Experiments were performed on alphabet sizes $2$, $5$, $10$ and $20$, and input words of lengths $1000$, $2000$, \ldots, $10000$. 
For each alphabet size and word length, $1000$ words were randomly generated.

In order to verify the correct execution of both algorithms, words have been generated in such a way that they have at least one non trivial Abelian period. Both algorithms have been compiled with the same options and run on the same machine under the same conditions.

\begin{figure}
\begin{center}
\large{(A) period 5}
\scalebox{.78}{
\begin{tabular}{c*{11}{c|}}
\cline{3-12}
&&\multicolumn{10}{|c|}{Word length}\\
\cline{3-12}
&&{\small 1000}&{\small 2000}&{\small 3000}&{\small 4000}&{\small 5000}&{\small 6000}&{\small 7000}&{\small 8000}&{\small 9000}&{\small 10000}\\
\hline
\multicolumn{1}{|c|}{\multirow{4}{.4cm}{\rotatebox{90}{~Alphabet~}}}&\multicolumn{1}{|c|}{2} & 
7.90 &
9.64&
10.51&
11.50 &
11.45&
11.42&
12.02&
11.64&
11.38&
11.62
\\
\multicolumn{1}{|c|}{} & \multicolumn{1}{|c|}{5} & 
7.38 &
9.03&
9.50 &
10.32&
10.26 &
10.32 &
10.54&
10.38 &
10.34&
10.46
\\
\multicolumn{1}{|c|}{} & \multicolumn{1}{|c|}{10} & 
7.04&
8.28 &
8.67 &
9.37 &
9.03 &
9.13 &
9.55&
9.47&
9.33&
9.47
\\
\multicolumn{1}{|c|}{} & \multicolumn{1}{|c|}{20} & 
6.13&
6.66&
6.93&
7.08&
7.14 &
7.16 &
7.31 &
7.30&
7.18 &
7.20
\\
\hline
\end{tabular}
}
~\\
\large{(B) period 20}
\scalebox{.78}{
\begin{tabular}{c*{11}{c|}}
\cline{3-12}
&&\multicolumn{10}{|c|}{Word length}\\
\cline{3-12}
&&{\small 1000}&{\small 2000}&{\small 3000}&{\small 4000}&{\small 5000}&{\small 6000}&{\small 7000}&{\small 8000}&{\small 9000}&{\small 10000}\\
\hline
\multicolumn{1}{|c|}{\multirow{4}{.4cm}{\rotatebox{90}{~Alphabet~}}}&\multicolumn{1}{|c|}{2} & 
8.88 &
10.72 &
11.06 &
11.88 &
11.67&
11.87 &
12.32 &
12.48&
12.32&
12.90
\\
\multicolumn{1}{|c|}{} & \multicolumn{1}{|c|}{5} & 
11.33&
13.98 &
15.47 &
15.99 &
15.91 &
16.70 &
16.71 &
16.98&
16.57 &
16.77
\\
\multicolumn{1}{|c|}{} & \multicolumn{1}{|c|}{10} & 
10.93 &
13.30 &
14.45&
14.95&
14.69&
15.57&
15.62 &
15.90 &
15.44&
16.03
\\
\multicolumn{1}{|c|}{} & \multicolumn{1}{|c|}{20} & 
8.81 &
11.72 &
12.46&
12.74&
12.46&
13.06 &
13.07 &
13.17 &
13.35&
13.59
\\
\hline
\end{tabular}
}
\end{center}
\caption{\label{fig-ratio1}Execution time ratios for LFAP over QLFAP algorithms. 
 (A) Ratios computed on random words with at least one non trivial Abelian period 5. (B) Ratios computed on random words with at least one non trivial Abelian period 20.}
\end{figure}

\begin{figure}
\begin{center}
\large{(A) period 5}
\scalebox{.78}{
\begin{tabular}{c*{11}{c|}}
\cline{3-12}
&&\multicolumn{10}{|c|}{Word length}\\
\cline{3-12}
&&{\small 10000}&{\small 20000}&{\small 30000}&{\small 40000}&{\small 50000}&{\small 60000}&{\small 70000}&{\small 80000}&{\small 90000}&{\small 100000}\\
\hline
\multicolumn{1}{|c|}{\multirow{4}{.4cm}{\rotatebox{90}{~Alphabet~}}}&\multicolumn{1}{|c|}{2} & 
11.62&
12.38 &
12.32 &
15.82 &
15.86 &
15.84 &
16.27&
16.65 &
15.94&
16.72
\\
\multicolumn{1}{|c|}{} & \multicolumn{1}{|c|}{5} & 
10.46&
10.85 &
11.25&
13.67 &
13.84 &
13.90&
14.34&
14.31&
13.98 &
14.46
\\
\multicolumn{1}{|c|}{} & \multicolumn{1}{|c|}{10} & 
9.47 &
10.07 &
10.30&
11.86 &
11.86&
11.86&
12.13&
12.08&
11.86&
12.20
\\
\multicolumn{1}{|c|}{} & \multicolumn{1}{|c|}{20} & 
7.20 &
7.65&
7.71 &
8.54&
8.57&
8.59 &
8.71 &
9.24&
8.75&
8.92
\\
\hline
\end{tabular}
}
~\\
\large{(B) period 20}
\scalebox{.78}{
\begin{tabular}{c*{11}{c|}}
\cline{3-12}
&&\multicolumn{10}{|c|}{Word length}\\
\cline{3-12}
&&{\small 10000}&{\small 20000}&{\small 30000}&{\small 40000}&{\small 50000}&{\small 60000}&{\small 70000}&{\small 80000}&{\small 90000}&{\small 100000}\\
\hline
\multicolumn{1}{|c|}{\multirow{4}{.4cm}{\rotatebox{90}{~Alphabet~}}}&\multicolumn{1}{|c|}{2} & 
12.90 &
13.43&
13.44 &
17.00&
17.23&
17.46 &
17.63&
17.21&
18.37 &
18.10
\\
\multicolumn{1}{|c|}{} & \multicolumn{1}{|c|}{5} & 
16.77&
17.80 &
18.45 &
24.27&
24.58&
24.69 &
26.55 &
27.07 &
27.20 &
28.42
\\
\multicolumn{1}{|c|}{} & \multicolumn{1}{|c|}{10} & 
16.03&
16.50 &
17.17&
21.80&
22.46&
22.85 &
24.47&
25.31 &
25.09&
25.45
\\
\multicolumn{1}{|c|}{} & \multicolumn{1}{|c|}{20} & 
13.59 &
13.73 &
14.37 &
17.41&
18.27 &
18.49 &
19.43 &
19.07 &
19.02&
19.33
\\
\hline
\end{tabular}
}
\end{center}
\caption{\label{fig-ratio2}Execution time ratios for LFAP over QLFAP algorithms. 
 (A) Ratios computed on random words with at least one non trivial Abelian period 5. (B) Ratios computed on random words with at least one non trivial Abelian period 20.}
\end{figure}

\begin{figure}
\begin{center}
\large{(A) period 5}
\scalebox{.78}{
\begin{tabular}{c*{8}{c|}}
\cline{3-8}
&&\multicolumn{6}{|c|}{Word length}\\
\cline{3-8}
&&{\small 50000}&{\small 100000}&{\small 200000}&{\small 400000}&{\small 800000}&{\small 1600000}\\
\hline
\multicolumn{1}{|c|}{\multirow{4}{.4cm}{\rotatebox{90}{~Alphabet~}}}&\multicolumn{1}{|c|}{2} & 
15.86&16.72&17.87&19.29&19.50&20.35
\\
\multicolumn{1}{|c|}{} & \multicolumn{1}{|c|}{5} & 
13.84&14.42&15.44&16.24&16.99&17.66
\\
\multicolumn{1}{|c|}{} & \multicolumn{1}{|c|}{10} & 
11.86&12.20&12.88&13.89&14.44&15.08
\\
\multicolumn{1}{|c|}{} & \multicolumn{1}{|c|}{20} & 
8.54&8.92&9.53&9.34&10.07&10.57
\\
\hline
\end{tabular}
}
~\\
\large{(B) period 20}
\scalebox{.78}{
\begin{tabular}{c*{8}{c|}}
\cline{3-8}
&&\multicolumn{6}{|c|}{Word length}\\
\cline{3-8}
&&{\small 50000}&{\small 100000}&{\small 200000}&{\small 400000}&{\small 800000}&{\small 1600000}\\
\hline
\multicolumn{1}{|c|}{\multirow{4}{.4cm}{\rotatebox{90}{~Alphabet~}}}&\multicolumn{1}{|c|}{2} & 
17.23&18.10&20.04&20.82&21.49&23.39
\\
\multicolumn{1}{|c|}{} & \multicolumn{1}{|c|}{5} & 
24.58&29.42&30.21&30.76&31.16&34.13
\\
\multicolumn{1}{|c|}{} & \multicolumn{1}{|c|}{10} & 
22.46&25.45&27.84&29.42&28.33&31.23
\\
\multicolumn{1}{|c|}{} & \multicolumn{1}{|c|}{20} & 
18.27&19.33&21.17&22.05&22.42&24.45
\\
\hline
\end{tabular}
}
\end{center}
\caption{\label{fig-ratio3}Execution time ratios for LFAP over QLFAP algorithms. 
 (A) Ratios computed on random words with at least one non trivial Abelian period 5. (B) Ratios computed on random words with at least one non trivial Abelian period 20.}
\end{figure}

The plots show that, on the range of tested values, the two algorithms
 have a linear behaviour.
The results show that our algorithm is significantly faster than 
 \textsc{LinearFullAbelianPeriods} algorithm. 
One can also observe that the ratio increases both with word length
 and with the length of the Abelian period (see \figurename~\ref{fig-ratio3}).
This may be due to the larger number of arrays needed by the  \textsc{LinearFullAbelianPeriods} algorithm. 

\begin{figure}
\centering{
  \subfloat[]{\label{fig:1000}
\includegraphics[width=5.5cm,angle=-90]{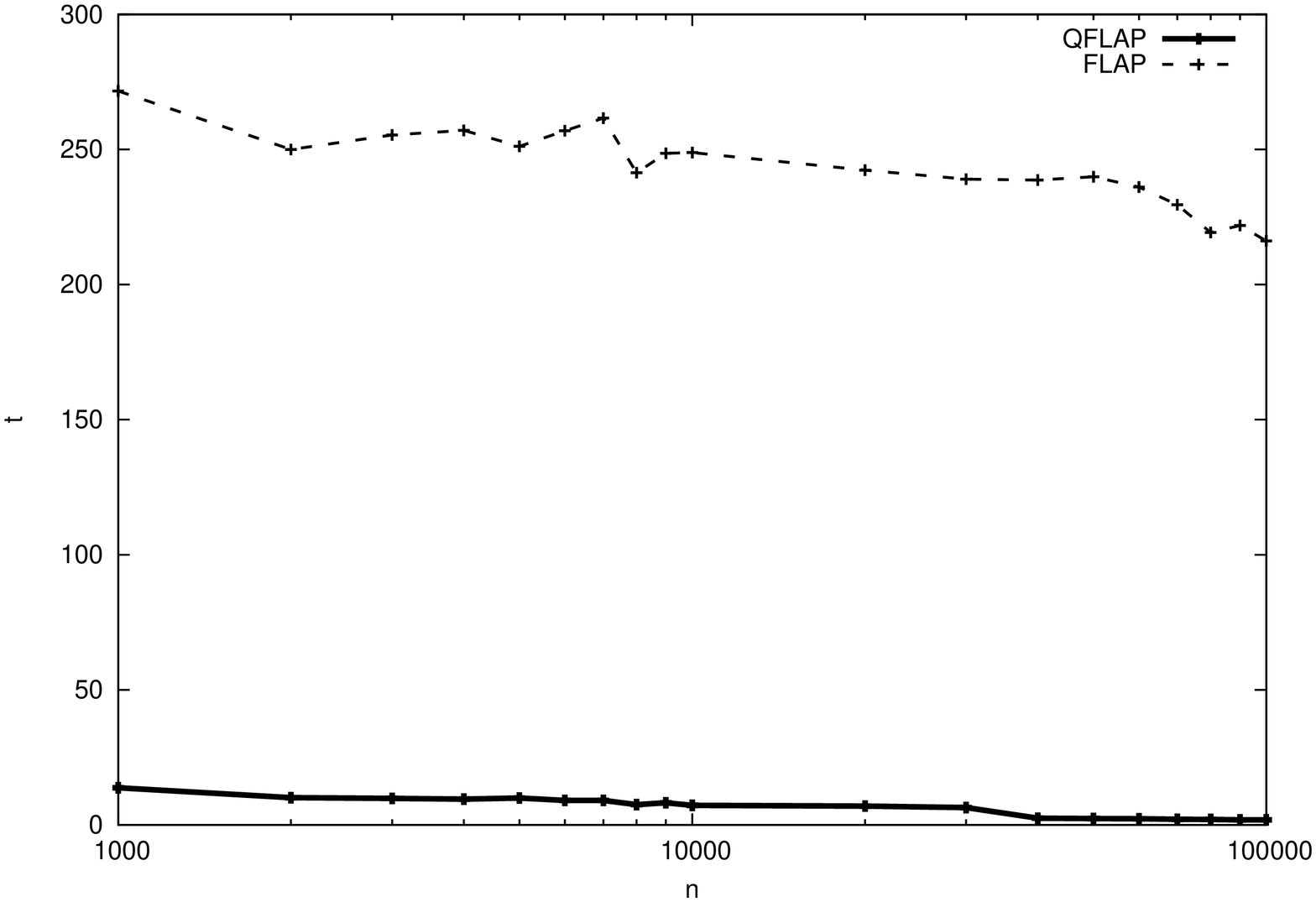}
}
}
\caption{
\label{fig-expe-ecoli}
Running times of the LFAP and QLFAP algorithms for
the \texttt{E.coli} file from the Canterbury Large Corpus.
}
\end{figure}

Similar results were obtained with the \texttt{E.coli} file from the Canterbury Large Corpus
 (\figurename s~\ref{fig-expe-ecoli} and \ref{fig-ratio-ecoli}).

\begin{figure}
\begin{center}
\begin{tabular}{|l|c|c|c|c|c|c|c|c|c|c|}
\hline
Word length & {\small 1000}&{\small 2000}&{\small 3000}&{\small 4000}&{\small 5000}&{\small 6000}&{\small 7000}&{\small 8000}&{\small 9000}&{\small 10000}\\
\hline
Ratio &
19.72 &
24.79&
26.04 &
26.95&
25.23&
28.31 &
28.79 &
31.93&
30.24&
31.51
\\
\hline
\end{tabular}

\begin{tabular}{|l|c|c|c|c|c|c|c|c|c|}
\hline
Word length &
{\small 20000}&{\small 30000}&{\small 40000}&{\small 50000}&{\small 60000}&{\small 70000}&{\small 80000}&{\small 90000}&{\small 100000}\\
\hline
Ratio &
34.75 &
37.10 &
95.35 &
101.73 &
102.57 &
107.82&
108.00&
116.06&
114.54
\\
\hline
\end{tabular}
\end{center}
\caption{\label{fig-ratio-ecoli}Execution time ratio between LFAP and QLFAP algorithms on
 words taken from the \texttt{E.coli} file from the Canterbury Large Corpus.
}
\end{figure}

The source codes of LFAP and QLFAP algorithms can be found at
 \url{bioinfo.univ-rouen.fr/qlfap}.

\section{Conclusions and perspectives}\label{sec-conc}

In this note we presented an algorithm for computing all the full Abelian periods of a word of length $n$ in time $O(n\log\log n)$ and space $O(n)$. We showed that this algorithm is very efficient in practice, even if its theoretical worst-case behavior is superlinear.

Our experiments show that it is between 6 to 19 times faster than the linear time algorithm of~\cite{KoRaRy13} and we can expect that the ratio will be even larger for longer strings.
Our method requires less space.
In addition the latter algorithm needs to compare real numbers, which can be tricky in practice, while
our method avoids this difficulty.
Furthermore, our algorithm is conceptually simple, and stems from the notion of scaled factorization of a word, which could be of some interest in other related problems on abelian combinatorics on words.

A possible direction of further investigation could be that of extending the same approach to the computation of all the abelian periods with empty head but non-empty tail, or even to compute the smallest of such abelian periods.

\section*{Acknowledgements}

The authors thank the anonymous reviewers that greatly improved a first version
 of this paper.

\end{document}